\documentclass[aps,prl,floatfix,twocolumn,superscriptaddress]{revtex4-2}

\usepackage{graphicx,hyperref,siunitx,titlesec,bm}
\usepackage{subfigure}
\usepackage[toc,page]{appendix}
\usepackage{float}

\hypersetup{
	colorlinks=true,
	citecolor=blue,
	linkcolor=blue,
	urlcolor=blue}

\usepackage{amsmath}
\usepackage{amsfonts}
\usepackage{amssymb}
\usepackage{braket}
\usepackage{bm}
\usepackage{times}
\usepackage{scalerel}

\usepackage{tikz}
\usetikzlibrary{svg.path}

\usepackage{upgreek}
\usepackage{algpseudocode}
\usepackage{algorithm}

\graphicspath{{figures/}}
\newcommand{\myref}[2][]{Fig.~\hyperref[#2]{\ref*{#2}#1}}
\newcommand{\Myref}[2][]{Figure~\hyperref[#2]{\ref*{#2}#1}}

\definecolor{orcidlogocol}{HTML}{A6CE39}
\tikzset{
	orcidlogo/.pic={
		\fill[orcidlogocol] svg{M256,128c0,70.7-57.3,128-128,128C57.3,256,0,198.7,0,128C0,57.3,57.3,0,128,0C198.7,0,256,57.3,256,128z};
		\fill[white] svg{M86.3,186.2H70.9V79.1h15.4v48.4V186.2z}
		svg{M108.9,79.1h41.6c39.6,0,57,28.3,57,53.6c0,27.5-21.5,53.6-56.8,53.6h-41.8V79.1z M124.3,172.4h24.5c34.9,0,42.9-26.5,42.9-39.7c0-21.5-13.7-39.7-43.7-39.7h-23.7V172.4z}
		svg{M88.7,56.8c0,5.5-4.5,10.1-10.1,10.1c-5.6,0-10.1-4.6-10.1-10.1c0-5.6,4.5-10.1,10.1-10.1C84.2,46.7,88.7,51.3,88.7,56.8z};
	}
}

\newcommand\orcidicon[1]{\href{https://orcid.org/#1}{\mbox{\scalerel*{
				\begin{tikzpicture}[yscale=-1,transform shape]
					\pic{orcidlogo};
				\end{tikzpicture}
			}{|}}}}

\begin{document}
	
	\title{Synchronization in rotating supersolids}
	
	\author{Elena Poli\,\orcidicon{0000-0002-1295-9097}}
	\thanks{These authors contributed equally to this work.}
	\affiliation{Universität Innsbruck, Fakultät für Mathematik, Informatik und Physik,
		Institut für Experimentalphysik, 6020 Innsbruck, Austria}
	
	\author{Andrea Litvinov\,\orcidicon{0009-0001-5332-1188}}
	\thanks{These authors contributed equally to this work.}
	\affiliation{Institut f\"{u}r Quantenoptik und Quanteninformation, \"Osterreichische Akademie der \\ Wissenschaften, Technikerstr. 21A, 6020 Innsbruck, Austria}
	
	\author{Eva Casotti\,\orcidicon{0000-0002-8340-1445}}
	\affiliation{Institut f\"{u}r Quantenoptik und Quanteninformation, \"Osterreichische Akademie der \\ Wissenschaften, Technikerstr. 21A, 6020 Innsbruck, Austria}
	\affiliation{Universität Innsbruck, Fakultät für Mathematik, Informatik und Physik,
		Institut für Experimentalphysik, 6020 Innsbruck, Austria}
	
	\author{Clemens Ulm\,\orcidicon{0009-0009-7589-9536}}
	\affiliation{Institut f\"{u}r Quantenoptik und Quanteninformation, \"Osterreichische Akademie der \\ Wissenschaften, Technikerstr. 21A, 6020 Innsbruck, Austria}
	\affiliation{Universität Innsbruck, Fakultät für Mathematik, Informatik und Physik,
		Institut für Experimentalphysik, 6020 Innsbruck, Austria}
	
	\author{Lauritz Klaus\,\orcidicon{0000-0002-6018-0811}}
	\affiliation{Institut f\"{u}r Quantenoptik und Quanteninformation, \"Osterreichische Akademie der \\ Wissenschaften, Technikerstr. 21A, 6020 Innsbruck, Austria}
	\affiliation{Universität Innsbruck, Fakultät für Mathematik, Informatik und Physik,
		Institut für Experimentalphysik, 6020 Innsbruck, Austria}

	\author{Manfred J. Mark\,\orcidicon{0000-0001-8157-4716}}
	\affiliation{Universität Innsbruck, Fakultät für Mathematik, Informatik und Physik,
		Institut für Experimentalphysik, 6020 Innsbruck, Austria}
	\affiliation{Institut f\"{u}r Quantenoptik und Quanteninformation, \"Osterreichische Akademie der \\ Wissenschaften, Technikerstr. 21A, 6020 Innsbruck, Austria}
	
	\author{Giacomo Lamporesi\,\orcidicon{0000-0002-3491-4738}}
	\affiliation{Pitaevskii BEC Center, CNR-INO and Dipartimento di Fisica, Universit\`a di Trento, 38123 Trento, Italy}
	
	\author{Thomas Bland\,\orcidicon{0000-0001-9852-0183}}
	\affiliation{Universität Innsbruck, Fakultät für Mathematik, Informatik und Physik,
		Institut für Experimentalphysik, 6020 Innsbruck, Austria}
	
	\author{Francesca Ferlaino\,\orcidicon{0000-0002-3020-6291}}
	\thanks{Correspondence should be addressed to: \mbox{\url{francesca.ferlaino@uibk.ac.at}}}
	\affiliation{Universität Innsbruck, Fakultät für Mathematik, Informatik und Physik,
		Institut für Experimentalphysik, 6020 Innsbruck, Austria}
	\affiliation{Institut f\"{u}r Quantenoptik und Quanteninformation, \"Osterreichische Akademie der \\ Wissenschaften, Technikerstr. 21A, 6020 Innsbruck, Austria}
	
	\begin{abstract} 
		Synchronization is ubiquitous in nature at various scales and fields. This phenomenon not only offers a window into the intrinsic harmony of complex systems, but also serves as a robust probe for many-body quantum systems. One such system is a supersolid: an exotic state that is simultaneously superfluid and solid.
		Here, we show that putting a supersolid under rotation leads to a synchronization of the crystal's motion to an external driving frequency triggered by quantum vortex nucleation, revealing the system's dual solid-superfluid response.
		Benchmarking the theoretical framework against experimental observations, we exploit this model as a novel method to investigate the critical frequency required for vortex nucleation. Our results underscore the utility of synchronization as a powerful probe for quantum systems.
	\end{abstract}
	
	\maketitle
	
	Synchronization is a fundamental process whereby two or more distinct oscillators, initially operating at different intrinsic frequencies, adjust their rhythms, eventually evolving to oscillate in unison\,\cite{Pikovsky2001sau}. Huygens' synchronization, named after its discoverer in the 17th century, is a remarkable example of this phenomenon\,\cite{huygens1893ecd}. He observed that two pendulum clocks, when attached to a common support, eventually synchronize their frequency. Huygens called this the ``sympathy of two clocks'', noting that the weak motion of the shared support enabled their synchronization. 
	This early observation laid the foundation for understanding synchronization as a coupling-driven adjustment of rhythms.
	Nowadays, synchronization has become a widely recognized phenomenon that manifests across a broad range of natural and engineered systems. For instance, it is at the foundation of biological phenomena, ranging from the synchronous variation of cell nuclei\,\cite{banfalvi2011ooc} and the firing of biological oscillators like heart cells\,\cite{mirollo1990sop} to the coordinated blinking of fireflies\,\cite{buck1966bos}.
	More recently, the concept of synchronization has expanded into the realm of quantum physics\,\cite{holmes2012som,manzano2013sqc,mari2013moq,ameri2015mia}, being proposed as a signature of quantum correlation and entanglement\,\cite{witthaut2017csi,roulet2018qsa} and as an important mechanism in preventing quantum many-body systems from dephasing\,\cite{qiu2015hsi}. Furthermore, synchronization phenomena include also rotators under the effect of external  driving\,\cite{Pikovsky2001sau}, leading to intriguing phenomena such as frequency locking in Josephson junctions\,\cite{Barone1982paa,Oppenlander1996npd}.
	
	The study of synchronization in coupled oscillators or rotators becomes particularly fascinating when these entities represent entangled subsystems within a single many-body quantum state or are linked to distinct spontaneously broken symmetries of the same system. In the latter case, critical questions emerge: How do the oscillatory dynamics associated with each broken symmetry behave
	under external driving? Can their motion uncover novel collective phenomena, like frequency entrainment, and offer deeper insights into the coexistence of these symmetries? 
	
	Supersolid states of matter offer a compelling example, in which two symmetries spontaneously break simultaneously \cite{Gross1957uto,Leggett1970cas,Boninsegni2012csw}. These are global gauge symmetry, responsible for macroscopic phase coherence, and translational symmetry, which establishes crystalline order within the system.
	Recently, many-body quantum states with supersolid properties have attracted a great interest in a broad range of low-energy systems, including ultracold atoms \,\cite{Leonard2017sfi,Li2017asp,Boettcher2019tsp,Tanzi2019ooa,Chomaz2019lla}, helium crystals\,\cite{Levitin2019efa,Nyeki2017isa}, superconductors\,\cite{liu2023pdw,hamidian2016doa,xiang2024gme}, and are even predicted in high-energy matter such as neutron stars\,\cite{chamel2012nci,Poli2023gir}.
	
	Among these diverse platforms, dipolar quantum gases set the paradigm\,\cite{Norcia2021dia,Chomaz2022dpa}. In such systems, the interplay between short-range contact interactions, characterized by a tunable $s$-wave scattering length $a_s$, long-range magnetic dipole-dipole interactions with a fixed dipolar length $a_\text{dd}$, and dipolar-enhanced quantum fluctuations leads to the emergence of supersolid ground states\,\cite{Chomaz2022dpa}. 
	What makes a supersolid intriguing in the context of synchronization is the coexistence of superfluid and solid nature -- each distinctly responding to external perturbation, while being described by a single macroscopic wave function. This duality raises key questions. For instance, how can irrotationality of the superfluid flow coexist with rigid body rotation of the solid in a supersolid system of indistinguishable particles?  Is there a ``clock sympathy'' between the superfluid and solid responses that enable synchronized motion, or do they operate independently in this unique quantum state? 
	
	\begin{figure*}[t!]
		\includegraphics[width=1.9\columnwidth]{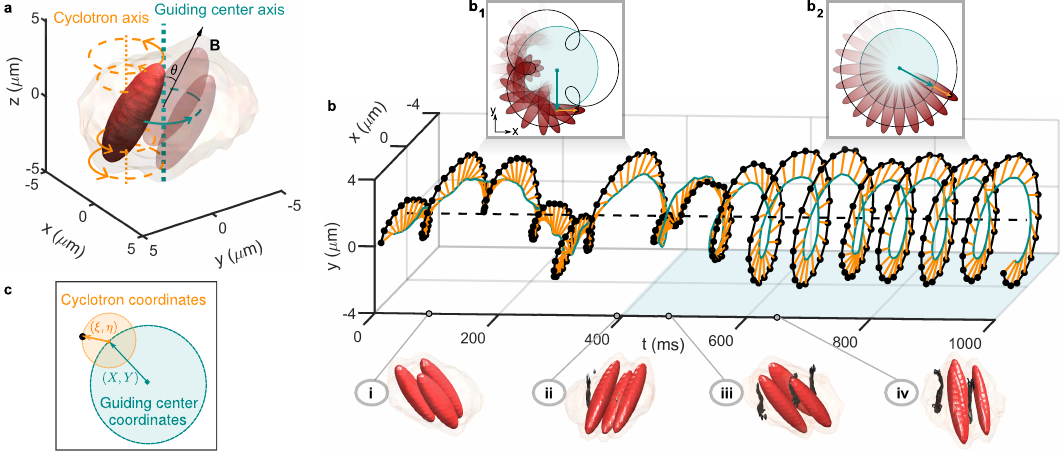}
		\caption{\textbf{Synchronization and concurrent vortex nucleation.} (a) Rotating supersolid from eGPE simulation, the isosurfaces are at $20\%$ (red) and $0.8\%$ (beige) of the maximum density, representing three droplets (one droplet highlighted in plain red color) and the halo, respectively. (b) In-plane trajectory of the droplet's tip (black) together with its decomposition in guiding center (green) and cyclotron (orange) motions. The light-blue shaded area highlights the time window in which a vortex is detected inside the system. The 3D isosurfaces (i)-(iv) correspond to different density frames during the synchronization process, with the black tubes corresponding to vortices. The insets show a schematic illustration of the droplet epitrochoidal (b$_1$) and circular (b$_2$) trajectory followed by the droplet. (c) Schematic representation of the decomposition in guiding center and cyclotron coordinates. The results are obtained for parameters: $a_\text{dd}\,{=}\,130.8\,a_0$, trap frequencies $[\omega_\perp,\omega_z] = 2\pi \times[50,95]$\,Hz, atom number $N=50000$, $a_s=95\,a_0$, magnetic field tilt angle $\theta = 30^\circ$, $\Omega = 2\pi\times 15$\,Hz, and dissipation constant $\gamma=0$.}
		\label{fig:fig1}
	\end{figure*}
	
	We address these questions through a combined theoretical and experimental investigation on the behavior of a rotating supersolid. Following theoretical predictions\,\cite{henkel2012svc,Roccuzzo2020ras,Gallemi2020qvi,Ancilotto2021vpi,casotti2024oov,Poli2023gir}, such systems have proven to be a fascinating playground to tackle general problems related to rotational flow\,\cite{Ancilotto2021vpi}, non-classical moment of inertia\,\cite{Roccuzzo2020ras,Poli2023gir}, and quantization of angular momentum\,\cite{Gallemi2020qvi} in the broad field of modulated superfluids\,\cite{bulgac2008ufs,henkel2012svc,casotti2024oov}. 
	Our starting points are the recent realization of circular supersolids exhibiting two-dimensional crystalline order\,\cite{Norcia2021tds,Bland2022tds} and the observation of vortex nucleation in this system\,\cite{casotti2024oov}. 
	In this study, we focus on the rotational dynamics by tracing the trajectory of the solid component and connect it to the vortex nucleation of the superfluid, presented in Fig.\,\ref{fig:fig1}. 
	
	Figure \ref{fig:fig1}(a) displays the calculated density isosurfaces for
	the supersolid ground state of a $^{164}$Dy dipolar quantum gas for our experimental parameters. 
	Our simulations employ the zero-temperature extended Gross–Pitaevskii equation (eGPE), incorporating quantum fluctuations\,\cite{Waechtler2016qfi,FerrierBarbut2016ooq,Chomaz2016qfd,Bisset2016gsp}; see Methods. This approach has been previously demonstrated as highly effective in capturing the complex properties of dipolar supersolid states\,\cite{Norcia2021tds,Bland2022tds,casotti2024oov}. 
	Note that, in displaying the isosurfaces, we have intentionally made the high-density crystalline peaks (hereafter referred to as droplets) and the low-density superfluid isosurface (halo) visually distinguishable. However, the particles are distributed continuously throughout the system, and there are not two separate components but a single, unified quantum state. 
	As shown in the figure, due to the magnetic nature of the dipolar interactions, the droplets align along the magnetic field axis, $\mathbf{B}$, whose axis is tilted relative to the vertical axis. 
	
	To impart angular momentum, we drive the system through a rotating magnetic field at frequency $\Omega$, a technique named magnetostirring\,\cite{Prasad2019vlf,Klaus2022oov,casotti2024oov}. 
	Figure\,\ref{fig:fig1}(b) shows the real-time evolution of the system responding to rotation.
	The full dynamics are captured through following the trajectory of the tip of a single droplet (black filled circle in (a)), as shown by the solid black line in (b). Surprisingly, after a few hundred milliseconds, we observe a drastic change in the droplet's rotational dynamics. Initially, the motion exhibits a double helicoidal behavior, as a consequence of precession and revolution proceeding at different frequencies. This type of trajectory draws an epitrochoidal path, as illustrated in the inset (b$_1$). On a longer timescale, the motion evolves into a circular trajectory (b$_2$), in which precession and revolution are frequency locked. When studying the phase pattern of the supersolid during the evolution, we remarkably observe a concurrency between the appearance of vortex nucleation and the abrupt trajectory change. 
	This behavior can be clearly seen by the 3D density isosurfaces (i)-(iv) where the vortex cores are visualized by black tubes (see Methods).
	
	To gain further insights, each droplet's response to the drive can be decomposed into two circular motions. The revolution around the trap center axis (guiding center axis) is described by the coordinates $(X,Y)$, and the precession around its own vertical axis (cyclotron axis) is captured by the coordinates $(\xi,\eta)$\,\cite{chen1984itp,fletcher2021gsi,caldara2023msv}; see Fig.\,\ref{fig:fig1}(c).
	
	\begin{figure}
		\includegraphics[width=0.46\textwidth]{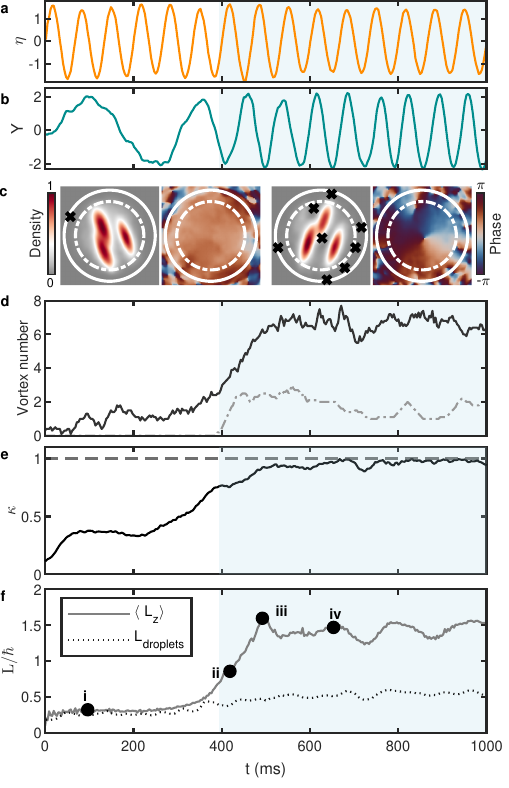}
		\caption{\textbf{Quantification of synchronization.} Time evolution of the cyclotron (a) and guiding center (b) coordinates. (c) Two exemplar frames showing the column density and the central phase slice of the rotating supersolid at $t=96.2$\,ms and $t=670.7$\,ms. The solid and dash-dotted circles mark the cutoff radii  $r^*=6\,\mu$m and $r^*=4.5\,\mu$m used to count the vortex number (d) time averaged over 35\,ms, with the same linestyle as the circles. (e) Frequency alignment $\kappa$, as defined in the main text. (f) Total angular momentum $\langle L_z\rangle$ and angular momentum of the droplets $L_\text{droplets}$, where (i)-(iv) refer to the ones of Fig.\,\ref{fig:fig1}(b). Across all subplots, the blue shaded region highlights when vortices enter within $r^*=4.5\,\mu$m. Parameters as in Fig.\,\ref{fig:fig1}.}
		\label{fig:fig2}
	\end{figure}
	
	These two components exhibit notably distinct behaviors, as shown in Fig.\,\ref{fig:fig2}(a) and (b). Throughout the time evolution, the cyclotron coordinates, describing precession, oscillate at a frequency, $\omega_c$, matching the external driving frequency $\Omega$ of the magnetic field $\mathbf{B}$. In contrast, the guiding center frequency, $\omega_g$, is initially significantly smaller than $\Omega$, then gradually increases, and eventually synchronizes with the external driving frequency; for the derivation of $\omega_c$ and $\omega_g$ see Methods. 
	We observe that synchronization occurs concurrently with nucleation of vortices in the system. Figure\,\ref{fig:fig2}(d) shows the number of vortices extracted within two different radii of interest, see Fig.\,\ref{fig:fig2}(c). Nucleation in the outer density region already promotes frequency locking between the cyclotron and guiding center motion, as evidenced by the increase in the parameter $\kappa = 1-\left|\frac{\omega_c-\omega_g}{\omega_c+\omega_g}\right|$, which quantifies the degree of synchronization (i.\,e.\, frequency alignment); see Fig\,\ref{fig:fig2}(e). Eventually, fully synchronous motion ($\kappa=1$) occurs as vortices approach the center of the supersolid.
	
\begin{figure}
		\includegraphics[width=0.46\textwidth]{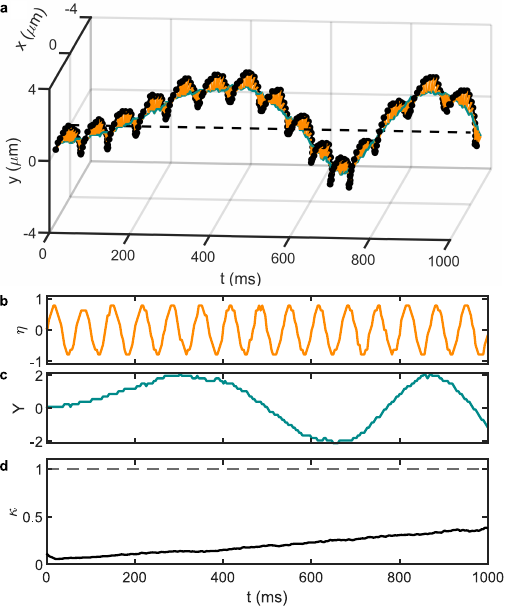}
		\caption{\textbf{Independent droplet regime.} (a) Time evolution of the droplet's edge trajectory and corresponding cyclotron (b) and guiding center coordinate (c). (d) Frequency alignment $\kappa$. All the results are obtained for the same parameters of Fig.\,\ref{fig:fig1}, except for $a_s=90\,a_0$.}
		\label{fig:fig3}
	\end{figure}

	When repeating the calculations of Fig.\,\ref{fig:fig2} for various $\Omega$ (Methods), we find that vortex-induced synchronization is a robust mechanism,  occurring across a wide range of $\Omega\geq\Omega^*$ values. Here, $\Omega^*$ is the critical frequency required for dynamical vortex nucleation~\cite{Fetter2009rtb,Gallemi2020qvi,casotti2024oov}. However, when rotating the supersolid at frequencies sufficiently high to have a ground state energetically supporting vortices, yet still below $\Omega^*$, our driven supersolid does not reach synchronization, i.\,e.\,equilibrium, in the considered timescale. Similar lack of equilibration is occurring in the isolated droplet regime; see Fig.\,\ref{fig:fig3} and later discussion.

	The concurrency between synchronization and vortex nucleation is clearly reflected by the behavior of the total angular momentum $\langle \hat{L}_z \rangle$, plotted in Fig.\,\ref{fig:fig2}(f). Initially, there is a low and constant value of $\langle \hat{L}_z \rangle$, (i). In this regime, $\langle \hat{L}_z \rangle \approx L_{\mathrm{droplets}} \approx I \times \Omega$, following a rigid-body rotation but with a non-classical moment of inertia $I$\,\cite{Leggett1970cas,Gallemi2020qvi}. Here, the response is dominated by the solid nature of the supersolid. 
	Around $t\sim\,400$ ms, $\langle \hat{L}_z \rangle$ rapidly increases when vortices move towards the center (ii), marking the onset of synchronization. Once the synchronization is complete ($\kappa\sim1$), the angular momentum stabilizes at a plateau (iii)-(iv), indicating that an equilibrium state has been reached in the rotating frame. 
	Here, small oscillations around the equilibrium value indicate variations in the number of vortices and their positions (iii)\,\cite{butts1999pso}, or excitations of the droplet lattice\,\cite{Poli2023gir}.
	
	These results reveal that the synchronization phenomenon can be seen as the process by which the supersolid acts as a driven rotator and dynamically approaches the ground state in the rotating frame, entrained by the rotating magnetic field~\cite{Pikovsky2001sau}. The solid part of the supersolid rigidly responds to rotation, whereas the superfluid part requires time for vortices to be nucleated and to provide the angular momentum to catch up to the driving frequency. By repeating similar real-time simulations but starting with a droplet crystal, with a negligible superfluid link connecting the droplets and no vortices, we observe the absence of synchronization ($\kappa<0.35$)  as shown in Fig.\,\ref{fig:fig3}. The gradual increase might be due to a residual halo connecting the droplets. 
	
	\begin{figure}[t!]
		\includegraphics[width=0.95\columnwidth]{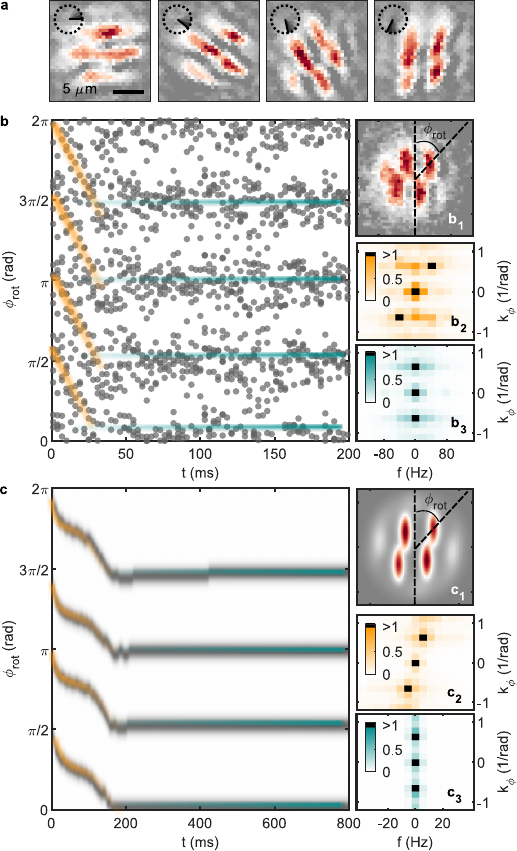}
		\caption{\textbf{Experimental observation of the synchronization process.} (a) Precession of the droplets in the lab frame at [1,13,25,37] ms, always aligned along the $\mathbf{B}$ field direction (black line). Angular position of the droplets (b$_1$, c$_1$) in the rotating frame as a function of time in experiment (b) and simulation (c). The orange and green lines are guides to the eye for the unsynchronized and synchronized cases, respectively.  (b$_2$, b$_3$) 2D Fourier transform of the experimental droplet angular position for early [$0,50$]\,ms and late [$60,110$]\,ms time intervals, respectively. (c$_2$, c$_3$) 2D Fourier transform of the theoretical droplet angular position for early [$0,200$]\,ms and late [$200,400$]\,ms time intervals, respectively. 
			The experimental data is taken for $\Omega=2\pi\times9$\,Hz,  trap frequencies $[\omega_\perp,\omega_z]=2\pi\times[50.5(6), 137(3)]\,\si{Hz}$, B=18.24(2)\,G, $N \approx 69000$.
			Theoretical simulations are done for $\Omega=2\pi\times9$\,Hz, $N=60000$, $a_s=90\,a_0$, trap frequencies $[\omega_\perp,\omega_z] = 2\pi\times[50,149]$\,Hz, dissipation parameter $\gamma=0.08$.}
		\label{fig:fig4}
	\end{figure}
	
	Building on our theoretical predictions, we explore potential synchronization phenomena experimentally. 
	In brief, our dipolar supersolid of $^{164}$Dy is produced via direct evaporative cooling\,\cite{Chomaz2019lla, Norcia2021tds}. We confine the system in a cylindrically symmetric optical dipole trap with harmonic frequencies $(\omega_\perp,\omega_z)$. During the last stages of evaporation, we tilt $\mathbf{B}$ by $\theta = 30^\circ$ and we produce a long-lived supersolid with four droplets. We then rotate the magnetic field with a constant angular velocity $\Omega=2\pi\times9$\,Hz for 300 ms following the magnetostirring protocol\,\cite{Klaus2022oov,casotti2024oov}. Finally, we track the position of the rotating droplets by taking destructive phase-contrast images along $z$ after $3\,\si{ms}$ of expansion, during which the $\textbf{B}$ is kept static and tilted. We observe the droplets precessing at constant frequency, always aligned along the direction of the $\textbf{B}$ field; see Fig.\,\ref{fig:fig4}(a). By fitting four Gaussian functions to the acquired column density profiles, we extract the center-of-mass position of each droplet and obtain their azimuthal angle $\phi_\text{rot} = \arctan(Y/X)$ to detect the guiding center motion, which is more convenient when plotting all droplet trajectories together. 
	We note that the initial tilt of the magnetic field breaks the cylindrical symmetry of the system, making the initial position of the four droplets repeatable over different experimental runs.
	
	Figure\,\Ref{fig:fig4} presents the experimental (b) and theoretical (c) results by plotting  $\phi_{\mathrm{rot}}$ of each of the four droplets. Visualizing the data in the rotating frame defined by $\Omega$ makes the effect of synchronization strikingly apparent. Constant $\phi_{\mathrm{rot}}$ means frequency locking with $\Omega$, whereas deviation from this behavior signalizes non-synchronous motion. 
	Both theory and experiment show that $\phi_{\mathrm{rot}}$ initially traces an oblique path, indicating that the droplets' center of mass is moving in the rotating frame. For later times, $\phi_{\mathrm{rot}}$ becomes constant: the signature of synchronization with the external driving frequency. Comparing two-dimensional Fourier transforms in $\phi_{\mathrm{rot}}$ and $t$ of the data for two selected time intervals, one at early times and the other at later times, further confirms this behavior. For each selected time interval, this gives three peaks, that reflect the presence or absence of periodic motion of droplets in the rotating frame.
	At early times, peaks appear at a finite frequency and produce a tilted pattern (b$_2$-c$_2$), indicating an asynchronous motion. At later times, instead, they align at zero frequency, the signature of synchronization (b$_3$-c$_3$).
	Both the data and their Fourier transform show an excellent agreement with the theory and confirm the experimental observation of synchronization. We note that, as is common in vortex studies, experimental noise and temperature effects lead to a faster vortex nucleation than the one predicted from mean-field theory \cite{Parker2005ead,Parker2005roa,Klaus2022oov,hernandez2024csf}. Moreover, the extra peaks in the experimental data arise due to shot-to-shot fluctuations in vortex nucleation time across different runs.

	\begin{figure}
		\includegraphics[width=0.9\columnwidth]{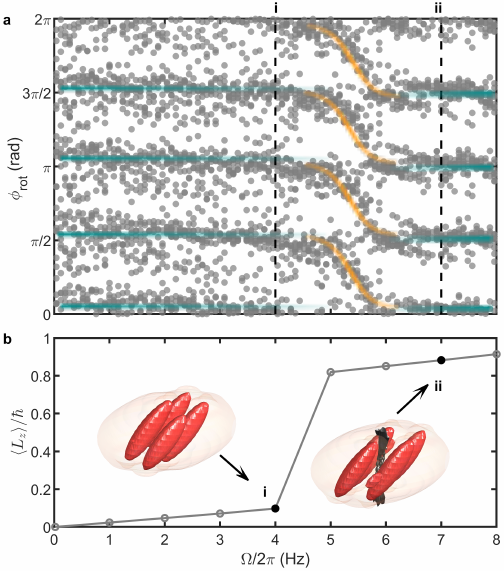}
		\caption{\textbf{Synchronization during a slow ramp of the driving rotation frequency.} (a)  $\phi_{\mathrm{rot}}$ measured at different points of the slow ramp of $\Omega$ from 0 to $2\pi\times$\SI{8}{Hz} in 200\,ms. The orange and green lines are a guide to the eye for the unsynchronized and synchronized cases, respectively. Experimental parameters: trap frequency $[\omega_\perp,\omega_z]=2\pi\times[50.3(4), 140.1(5)]\,\si{Hz}$, $N\approx69000$, B=18.30(2)\,G. (b) $\langle \hat{L}_z\rangle$ for ground states in the rotating frame varying $\Omega$ together with exemplary density isosurfaces. Simulation parameters: $N=70000$, $a_s=92\,a_0$, trap frequencies $2\pi\times[50,140]$\,Hz.}
		\label{fig:fig5}
	\end{figure}
	
	As the theory pinpoints a systematic correlation between vortex entering the system and the synchronization of the droplet motion, we can use the latter mechanism to further extract information on the vortex nucleation in the supersolid. Particularly interesting is the regime of low rotation frequencies and the quest of the minimal $\Omega$ for which a vortex is energetically stable. 
	To address this point, we now develop a different protocol. Instead of driving rotations at constant $\Omega$, we implement a scheme, in which $\Omega$ is slowly increased linearly from zero to $\Omega=2\pi\times8$\,Hz over 200\,ms. 
	During the initial part of the slow ramp, the system remains synchronized since it adiabatically follows its ground state in the rotating frame. This is shown in Fig.\,\Ref{fig:fig5} by the constant $\phi_{\mathrm{rot}}$ (straight green lines). Around $\Omega\sim 2\pi\times5$\,Hz the trajectory of the droplets exhibits a sudden change. Here, the system desynchronizes (orange lines) since it has to adjust to the new ground state in the rotating frame, now possessing a vortex.   
	Around $\Omega\sim 2\pi\times6$\,Hz, the system restores the equilibrium in the rotating frame and the synchronization condition, with the $\phi_{\mathrm{rot}}$ forming a straight horizontal path again, suggesting that the first vortex has entered. Thus, from this dynamics, we can infer an upper limit for the energetic critical rotation frequency for vortex nucleation to be $\Omega_c\le 2\pi\times6$\,Hz. This value is in excellent agreement with the predicted critical frequency, obtained from calculations of the supersolid ground-state in the rotating frame, as shown in Fig.\,\Ref{fig:fig5}(b). This result confirms that supersolids provide a good platform for dynamically exploring low-frequency vortex nucleation. The corresponding angular momentum exhibits a sudden jump around $\Omega_c\sim2\pi\times5$\,Hz when the ground state undergoes a transition from the zero- to the one-vortex state. The amplitude of the jump quantifies the angular momentum carried by the vortex at the center of the supersolid, which is equal to $0.72\hbar$. As pointed out in Ref.\,\cite{Gallemi2020qvi}, the sub-unity value is directly connected to the reduced superfluid fraction characteristic of modulated superfluids. The nucleation of a vortex at this low frequency is concurrent with a dynamical instability of the superfluid quadrupole mode, so far not observed \cite{casotti2024oov}.

	In conclusion, we have demonstrated that synchronization phenomena arise in rotating dipolar supersolids driven by an external magnetic field. We find that the synchronization process, driven by vortex nucleation, 
	reflects the system’s approach towards equilibrium and reveals the solid-superfluid dual nature of the supersolid. By decomposing the droplet motion into cyclotron and guiding center coordinates, we differentiate between precession around the cyclotron axis, which consistently synchronizes with the driving frequency, from global revolution.  At large rotation frequencies, the latter initially displays asynchronous motion. Here, while the supersolid's ground state in the rotating frame contains vortices, the driven supersolid begins out of equilibrium and is vortex-free. As it converges toward its ground state over time, the system only synchronizes when vortices enter. Furthermore, our analysis of the supersolid’s angular momentum confirms that synchronization arises from the delayed superfluid response, in contrast to the immediate solid-like response. Finally, we identify the synchronization as a novel diagnostic tool to measure the critical rotation frequency required for vortex nucleation.
	Future studies could use desynchronization and resynchronization during the slow-down of the driving frequency as a probe to understand vortex emission dynamics in analog to glitches observed in neutron stars \cite{Poli2023gir,Bland2024epg}.

	\noindent {\bf Acknowledgements:} 
	We acknowledge useful discussions with Zoran Hadzibabic, Massimo Mannarelli, Rosario Fazio, Russell Bisset, and Hannah Geiger. This work was supported by the European Research Council through the Advanced Grant DyMETEr (No.\,101054500) Grant DOI 10.3030/101054500, a joint-project grant from the Austrian Science Fund FWF (No.\,I-4426) Grant DOI 10.55776/I4426, a NextGeneration EU grant AQuSIM by the Austrian Research Promotion Agency FFG (No.\,FO999896041), the Austrian Science Fund (FWF) Grant DOI 10.55776/PAT1597224 and the Austrian Science Fund (FWF) Cluster of Excellence QuantA Grant DOI 10.55776/COE1. E.\,P.\,acknowledges support by the Austrian Science Fund (FWF) within the DK-ALM (No.\,W1259-N27) Grant DOI 10.55776/W1259. T.\,B.\,acknowledges financial support through an ESQ Discovery grant by the Austrian Academy of Sciences. A.\,L.\,acknowledges financial support through the Disruptive Innovation - Early Career Seed Money grant by the Austrian Science Fund FWF and Austrian Academy of Science ÖAW. G.\,L.\, acknowledges Provincia Autonoma di Trento for financial support.
	
	\noindent {\bf Author contributions} \\
	\noindent  A.L., E.C., C.U., L.K., M.J.M., G.L., and F.F. performed the experimental work and data analysis. E.P. and T.B. performed the theoretical work. All authors contributed to the interpretation of the results and the preparation of the manuscript.\\
	\noindent {\bf Data and materials availability:}
	Data pertaining to this work can be found on Zenodo \url{ https://doi.org/10.5281/zenodo.14193459}.\\
	\noindent {\bf Code availability:}
	The codes that support the findings of this study are available from the corresponding author upon reasonable request.

%

	\appendix

	\clearpage
	\begin{center}
		\textbf{\large Methods}
	\end{center}
	
	\setcounter{equation}{0}
	\setcounter{figure}{0}
	\setcounter{table}{0}
	\setcounter{page}{1}
	\makeatletter
	\renewcommand{\theequation}{S\arabic{equation}}
	\renewcommand{\figurename}{Extended Data Fig.}
	
	\section{Theoretical description}\label{formalism} 
	We study the ground state and dynamics of a supersolid state using an extended Gross-Pitaevskii formalism. We consider a supersolid made of $^{164}$Dy dipolar atoms of mass $m$ trapped in a cylindrically symmetric harmonic potential $V(x,y,z)=\frac{1}{2} m\left[\omega_\perp^2(x^2 + y^2) + \omega_z^2z^2\right]$, where $\omega_\perp$ ($\omega_z$) is the radial (axial) trap frequency. At zero temperature, the inter-particle interaction is described by the pseudo-potential
	\begin{align}
		U(\mathbf{r},t)=\frac{4\pi \hbar^2 a_{\mathrm{s}}}{m} \delta(\mathbf{r})+\frac{3 \hbar^2 a_{\mathrm{dd}}}{m} \frac{1-3(\hat{\mathbf{e}}(t) \cdot \mathbf{r})^2}{r^3}\,.
	\end{align} 
	The first term represents the short-range repulsive contact interaction, characterized by the s-wave scattering length $a_s$. This parameter can be experimentally adjusted via Feshbach tuning~\cite{Chin2010fri}. The second term represents the long-range anisotropic dipole-dipole interaction. For $^{164}$Dy atoms, $a_\text{dd}=130.8\,a_0$. For the experimentally relevant trap geometries and atom numbers, we find two-dimensional supersolid ground states for  $\epsilon_\text{dd}\,{=}\,a_\text{dd}/a_s\,{\gtrsim}\,1.3$. The unit vector ${\hat{\mathbf{e}}(t)=(\sin \theta \cos \varphi(t), \sin \theta \sin \varphi(t), \cos \theta)}$ indicates the polarization direction of the dipoles, which is set by the external magnetic field $\mathbf{B}$. In our study, $\mathbf{B}$ has a fixed angle of $\theta=30^\circ$ with respect to the vertical $z-$axis. The time-dependent azimuthal angle is $\varphi(t)=\int_0^t \mathrm{\,d} t^{\prime} \Omega\left(t^{\prime}\right)$, where $\Omega(t)$ is the angular velocity at time $t$. 
	
	Under this formalism, the extended Gross-Pitaevskii equation (eGPE) reads\,\cite{Waechtler2016qfi,Bisset2016gsp}
	\begin{equation}\label{epge}
		i \hbar \frac{\partial \Psi(\mathbf{r}, t)}{\partial t}=(\alpha-i \gamma)\mathcal{L}_{\mathrm{GP}}\left[\Psi(\mathbf{r}, t)\right] \Psi(\mathbf{r}, t)\,,
	\end{equation}
	where $\mathcal{L}_{\mathrm{GP}}$ is the Gross-Pitaevskii operator defined as
	\begin{equation}\label{epge-op}
		\begin{aligned}
			\mathcal{L}_{\mathrm{GP}}\left[\Psi(\mathbf{r}, t)\right] = & {\left[-\frac{\hbar^2 \nabla^2}{2 m}\right.} + V(x,y,z) \\
			& +\int \text{d}^3\mathbf{r}^{\prime}\, U\left(\mathbf{r}-\mathbf{r}^{\prime},t\right)\left|\Psi\left(\mathbf{r}^{\prime}, t\right)\right|^2  \\
			& +\gamma_{\mathrm{QF}}|\Psi(\mathbf{r}, t)|^3\biggr]\,,
		\end{aligned}
	\end{equation}
	with $\Psi(\mathbf{r}, t)$ being the condensate wave function. The last term of $\mathcal{L}_{\mathrm{GP}}$ represents the Lee-Huang-Yang correction describing quantum fluctuations \cite{Lee1957eae,Lima2011qfi}, with coefficient
	$\gamma_{\mathrm{QF}}=\frac{128 \hbar^2}{3 m} \sqrt{a_{\mathrm{s}}^5} \operatorname{Re}\left\{\mathcal{Q}_5\left(\epsilon_{\mathrm{dd}}\right)\right\}$, where $\mathcal{Q}_5\left(\varepsilon_{\mathrm{dd}}\right)=\int_0^1 d u\left(1-\varepsilon_{\mathrm{dd}}+3 u^2 \varepsilon_{\mathrm{dd}}\right)^{5 / 2}$. This term is necessary for the stabilization of the supersolid state against collapse \cite{Bisset2016gsp}. 
	
	In Eq.\,\eqref{epge} the parameters $\alpha$ and $\gamma$ determine the type of evolution. Imaginary time evolution $\left\{\alpha=0, \gamma = 1\right\}$ is used to find the ground state of the system. Real-time evolution corresponding to $\left\{\alpha=1, \gamma = 0\right\}$ is used in the study of synchronization dynamics in Figs.\,\ref{fig:fig1}, \ref{fig:fig2}, and \ref{fig:fig3}. Complex-time evolution with $\left\{\alpha=1, \gamma \neq 0\right\}$ has become a toy model for thermal dissipation that qualitatively captures very well the vortex dynamics, accelerating the nucleation process\,\cite{Pitaevskii1959pto,Choi1998pdi,Makoto2002vlf,Penckwitt2002nga,proukakis2024scs}. This approach provides a closer match to the experimental results shown in Fig.\,\ref{fig:fig4}. In practice, this technique can be interpreted as an hybrid of real-time and imaginary-time evolution. At each time step, the system undergoes real-time evolution in response to external perturbations, while simultaneously relaxing toward the stationary state under those conditions through imaginary-time.
	However, accounting for dissipation through a single constant parameter $\gamma$ is a simplified approach, whereas in experiments the situation is more complicated, e.\,g.\,the spatial dependence of the thermal cloud \cite{Blakie2008das}. Indeed, the observed dynamics are typically faster than the one in the simulation, and may call for more sophisticated finite temperature theories to reconcile this difference \cite{proukakis2024scs}.
	
	It is important to compare the results of the real-time evolution with the actual ground-state, towards which the system is expected to evolve.  We calculate the ground state in the rotating frame. The effective Hamiltonian in this frame includes a term $-\Omega \hat{L}_z$ in Eq.\,\eqref{epge-op}, where $\hat{L}_z=x\hat{p}_y - y\hat{p}_x$ is the angular momentum operator. We use this approach to generate Fig.\,\ref{fig:fig5}(b). 
	
	Both in the real (dynamics) and imaginary (ground-state) time evolution, we identify the presence of vortices inside the system by detecting $2\pi$ phase windings of the wave function within a radius $r^*$ centered at the origin. In Fig.\,\ref{fig:fig1} we use $r^* = 4.5\,\mu$m. To visualize the vortex tubes in Fig.\,\ref{fig:fig1} and Fig.\,\ref{fig:fig4}, we plot isosurfaces of the velocity field.
	
	\section{Coordinates decomposition}
\begin{figure}[h!]
		\includegraphics[width=0.5\columnwidth]{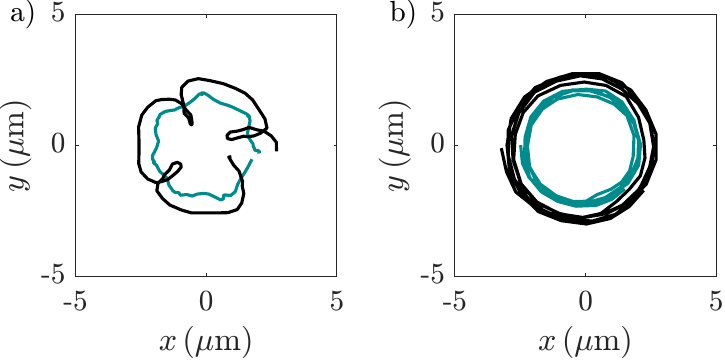}
		\caption{Trajectory of the droplet's tip (black line) and center of mass (green line) for the (a) unsynchronized and (b) synchronized state. For the unsynchronized state we show the time dynamics for $t<300$\,ms, and for the synchronized state we show the time dynamics for $t>800$\,ms. The other parameters are the same as Fig.\,\ref{fig:fig1} in the main text.}
		\label{fig:traj}
	\end{figure}

	To reveal the dual nature of the system's response, we decompose the droplet trajectory into its center-of-mass motion (guiding center coordinate) and precession motion (cyclotron coordinate). These two sets of coordinates are extensively used in e.\,g.\,plasma orbit theory \cite{chen1984itp} and, more recently, to study rotating BECs \cite{fletcher2021gsi,caldara2023msv}. We extract these coordinates by identifying the position of the density maxima in two distinct $z$-slices of the 3D wavefunction. First, we use the density slice at $z = 0$ to determine the position of the center of mass of each droplet relative to the origin, which we associate with the guiding center coordinates $(X, Y)$. Next, the density slice at $z = 2.5\,\mu$m is used to identify the position of the edge of each droplet, denoted as $(x_d, y_d)$. The cyclotron vector $(\xi, \eta)$ is then obtained by subtracting these two quantities: $(\xi, \eta) = (x_d, y_d) - (X, Y)$ (see Fig.\,\ref{fig:fig1}(c) in the main text). The choice of the $z$-slice for $z \neq 0$ influences the magnitude of the cyclotron vector but does not affect its orbital frequency. Importantly, our results remain robust regardless of the chosen $z$-slice and the choice of the droplet. In Extended Data Fig.\,\ref{fig:traj} we illustrate the in-plane projected position of the droplet's tip (black line) and droplet's center of mass (green line) for the (a) unsynchronized and (b) synchronized state, highlighting the analogy to the epitrochoidal and circular orbits shown as a sketch in the insets (b$_1$-b$_2$) of Fig.\,\ref{fig:fig1}. 
	For the calculation of the frequency alignment $\kappa$, we extract the instantaneous frequencies of the guiding center vector, $\omega_g$, and of the cyclotron vector, $\omega_c$. We compute those quantities by first extracting the angle of the guiding center and cyclotron vector at time $t$ from the respective coordinates, namely $\varphi_{g} = \arctan{(Y/X)}$ and $\varphi_{c} = \arctan{(\eta/\xi)}$. We then calculate the time derivatives $\omega_g = \Delta\varphi_{g}/\Delta t$ and $\omega_c = \Delta\varphi_{c}/\Delta t$, with $\Delta t=3.5$\,ms.

	\section{Droplet's angular momentum calculation}\label{angular_momentum} 
	The guiding center and cyclotron coordinates provide a comprehensive framework to describe the motion of the droplets during the synchronization process, offering a tool to estimate the droplet's angular momentum. In general, the total angular momentum $\langle\hat{L}_z\rangle$ of a rotating supersolid can be decomposed into two contributions, $L_{\mathrm{s}}$ and $L_{\mathrm{vort}}$ \cite{Roccuzzo2020ras,Gallemi2020qvi,Poli2023gir}. Here, $L_\mathrm{s}$ represents the angular momentum associated with the solid response, reflecting the system's non-zero moment of inertia, whereas $L_{\mathrm{vort}}$ is the superfluid angular momentum stored in the form of quantized vortices. 
	We estimate $L_{\mathrm{s}}$ by summing the contributions from both the cyclotron motion and the guiding center motion of the droplets, giving the droplet angular momentum 
	\begin{align}
		L_{\mathrm{droplets}} &= L_{\mathrm{guid}} + L_{\mathrm{cycl}} \nonumber \\ &=I_0\,\omega_{\mathrm{guid}} + \sum_{i=1}^{N_d} I_{\mathrm{droplets}}^{i}\,\omega_{\mathrm{cycl}}\,,
	\end{align}
	where $N_d$ is the number of droplets in the supersolid state.
	Since these motions occur around different axes, we compute two distinct moments of inertia. The moment of inertia $I_{\mathrm{droplets}}^{i}$ for the $i$-th droplet, rotating around its own axis, is calculated through\,\cite{FerrierBarbut2018smo} 
	\begin{equation}
		I_{\mathrm{droplets}}^{i}=\frac{1}{2} m N_i \frac{\left(\sigma_y^2-\sigma_x^2\right)^2}{\sigma_x^2+\sigma_y^2}\,,
	\end{equation}
	where $\sigma_x$ and $\sigma_y$ are the droplet’s widths, obtained by fitting the column density with a Gaussian, and $N_i$ is the estimated number of atoms in the droplet. The moment of inertia for the supersolid as a whole, rotating around the origin, is computed with 
	\begin{equation}
		I_0=m \frac{\left\langle y^2- x^2\right\rangle^2}{\left\langle y^2 + x^2\right\rangle}\,,
	\end{equation}
	where $\langle\cdot\rangle$ is the expectation value calculated for the initial total wave function, with the reference frame centered at $(0,0)$. Previous work has shown that this estimate deviates from the true moment of inertia by approximately $5\%$\,\cite{Roccuzzo2020ras}. 
	The evolution of $L_{\mathrm{droplets}}$ is shown in Fig.\,\ref{fig:fig2}(e).

	\section{Synchronization for different scattering lengths}
\begin{figure}[h!]
		\includegraphics[width=0.7\columnwidth]{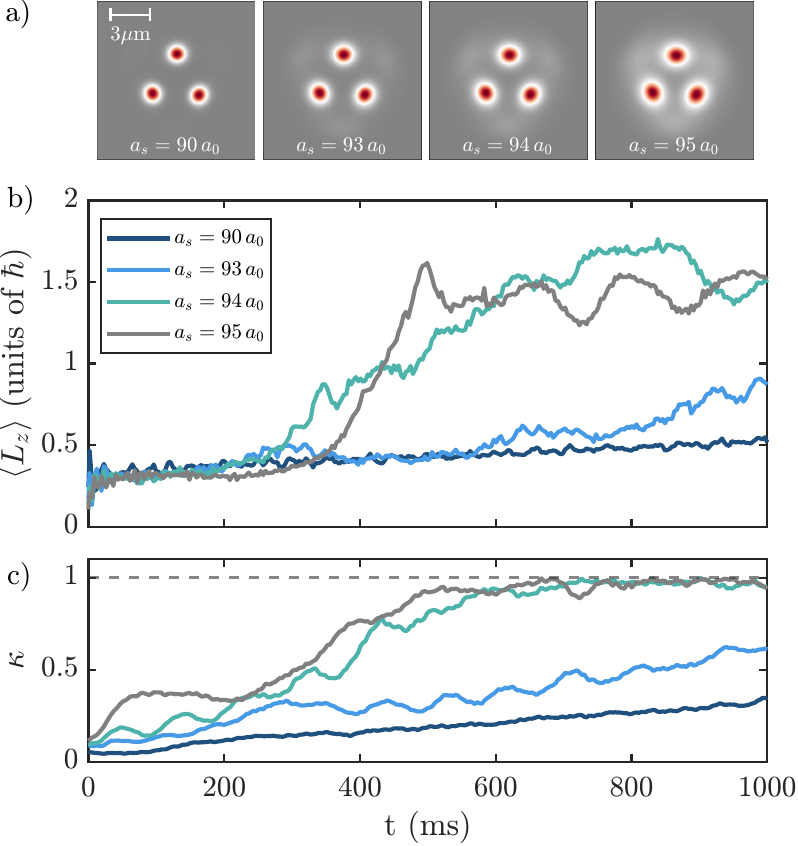}
		\caption{Synchronization process for different values of $a_s$. (a) Ground state density distributions with $\theta=0^\circ$, (b) angular momentum $\langle\hat{L}_z\rangle$ during the time evolution with $\theta=30^\circ$ and (c) frequency alignment $\kappa$. The other parameters are the same as Fig.\,\ref{fig:fig2}.}
		\label{fig:Lz_diffas}
	\end{figure}

	The synchronization process only occurs in the supersolid phase. This becomes clear when we study the angular momentum and frequency alignment $\kappa$ for different values of $a_s$. Extended data Fig.\,\ref{fig:Lz_diffas} shows these two quantities for initial states at different $a_s$, spanning from $a_s = 95\,a_0$, where the supersolid has strong superfluid connection between droplets, to $a_s = 90\,a_0$, in the independent droplet regime. In this data, we magnetostir at a constant frequency $\Omega = 2\pi\times15$ Hz. In the supersolid regime, the angular momentum exhibits a behavior similar to Fig.\,\ref{fig:fig2}(e) in the main text: it is initially constant, before suddenly increasing and ultimately stabilizing at a plateau when vortices move to the center. For this dynamical protocol at constant $\Omega$, the onset of vortex nucleation--marking the beginning of synchronization--reflects low frequency quadrupole mode resonances, which act as a seed for vortex nucleation, as discussed in Ref.\,\cite{casotti2024oov}. In contrast, in the independent droplet regime the angular momentum increases gradually without any sharp rise, indicating that vortices do not enter and synchronization is not occurring in an experimentally feasible timescale.
	
	To quantitatively relate the results across different scattering lengths, we compare the frequency alignment $\kappa$. In Extended data Fig.\,\ref{fig:Lz_diffas}(c) we show that synchronization is only achieved when the initial state is a supersolid. We interpret these results as follows: when the density coupling between droplets is negligible, this dynamical protocol is insufficient to bring the system to its equilibrium configuration in the rotating frame--where all droplets rotate at the same frequency--and, thus, synchronization fails to occur within an experimentally accessible timescale. We hypothesize that synchronization may never occur in the independent droplet regime in an experiment. In the theory, we use a single valued wavefunction that is forced to have a residual coupling between the droplets, eventually converging to a stationary solution to the underlying equation. Nevertheless, three-body losses prevent the experimental study of droplets in the independent droplet regime in the considered geometry.
	
	\section{Synchronization for different $\Omega$}
\begin{figure}[h!]
		\includegraphics[width=0.7\columnwidth]{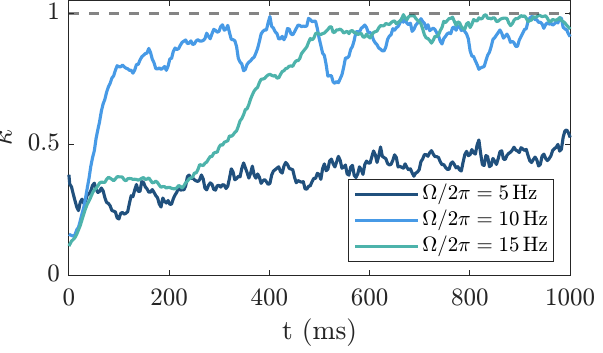}
		\caption{Frequency alignment $\kappa$ for different values of rotation frequency $\Omega$. The other simulation parameters are the same as Fig.\,\ref{fig:fig2}.}
		\label{fig:orderparam}
	\end{figure}

	It is interesting to consider the synchronization process for different fixed rotation frequencies. In Extended data Fig.\,\ref{fig:orderparam} we extend the synchronization analysis to three different driving frequencies $\Omega$. To compare the results, we use $\kappa$ as defined earlier. For $\Omega=2\pi\times5\,$Hz, $\kappa$ never reaches 1, meaning that synchronization does not occur within $1\,\si{s}$. However, with $\Omega=2\pi\times10\,$Hz, $\kappa$ grows faster than the one for $\Omega=2\pi\times15\,$Hz. As mentioned in the previous section, this earlier onset of vortex nucleation could be attributed to a resonance with a low-frequency quadrupole mode, which facilitates vortex formation, as suggested in Ref.\,\cite{casotti2024oov}. 
	In this analysis, we focused mainly on low rotation frequencies to isolate the role of individual vortices on the system's dynamics and to capture a few hundred milliseconds of pre-synchronization behavior. However, we expect similar dynamics to occur at higher rotation frequencies, albeit on faster timescales\,\cite{casotti2024oov}. 
	
\begin{figure}
		\centering
		\includegraphics[width=\linewidth]{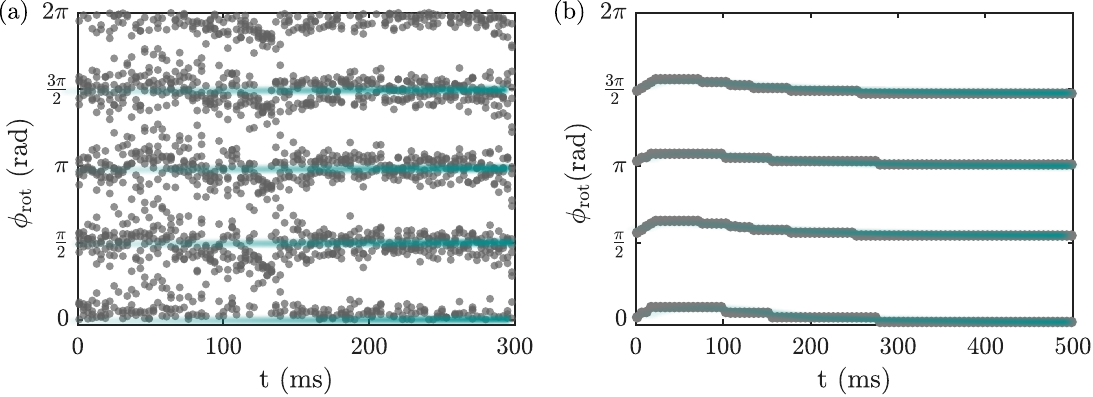}
		\caption{Angular position of the droplets in the rotating frame in the (a) experiment and (b) theory, when rotating at very low frequencies ($\Omega/2\pi = 3$\,Hz). The other parameters are the same as Fig.\,4 of the main text.}
		\label{fig:synchroNoVortex}
	\end{figure}

	Finally, we also show the experimental and theoretical rotation dynamics of a system whose ground state does not possess a vortex, see Extended data Fig.\,\ref{fig:synchroNoVortex}. This occurs for low rotation frequencies ($\Omega/2\pi = 3$\,Hz), for which the system is following the ground state and, thus, is always synchronized. At very short time dynamics, we observe a small motion in the rotating frame due to the excitation protocol with a sudden jump from 0 to 3 Hz, as expected.
	
	\section{Synchronization in a three droplet supersolid}\label{exp}
\begin{figure}
		\centering
		\includegraphics[width=0.7\linewidth]{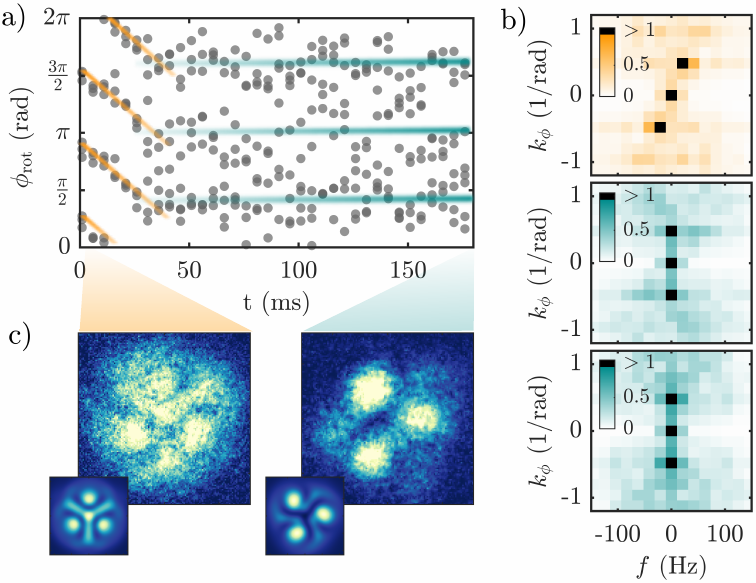}
		\caption{Synchronization process for a supersolid with three droplets. (a) Angular position of the droplets in the rotating frame. (b) 2D Fourier transform of the experimental droplet angular position for different time intervals. From top to bottom, the considered time intervals are [$0,50$]\,ms, [$20,70$]\,ms and [$40,90$]\,ms. (c) Exemplary absorption images of the interference pattern after 36\,ms expansion, in absence (left, $\Omega=0$\,Hz) and in presence (right, $\Omega=10$\,Hz after 189\,ms of rotation) of a vortex. The insets show the corresponding theoretical interference pattern obtained from real-time expansions. Parameters: $N =$ 35000, $[\omega_r,\omega_z] = [50.0(4),113(2)]$\,Hz, B=18.24(2)\,G, $a_s = 92.5a_0$.}
		\label{fig:synchro3droplets}
	\end{figure}

	In Extended Data Fig.\,\ref{fig:synchro3droplets} we present the experimental results of synchronization for a different supersolid state composed of three droplets, obtained for a different atom number and trap geometry~\cite{Poli2021msi}. The three droplet supersolid is the initial state studied in Ref.~\cite{casotti2024oov}, where the presence of vortices has been detected using multiple techniques. In Extended Data Fig.\,\ref{fig:synchro3droplets}, we show for this exact state both synchronization from in-situ images (as in Fig.\,4 of the main text) and time-of-flight interferometric measurements. This interferometric technique for vortex detection is in agreement with the numerical simulations and it has been shown to be robust and repeatable over different experimental shots~\cite{casotti2024oov}.

	\section{Experimental analysis}\label{exp}
\begin{figure}[h!]
		\includegraphics[width=1\columnwidth]{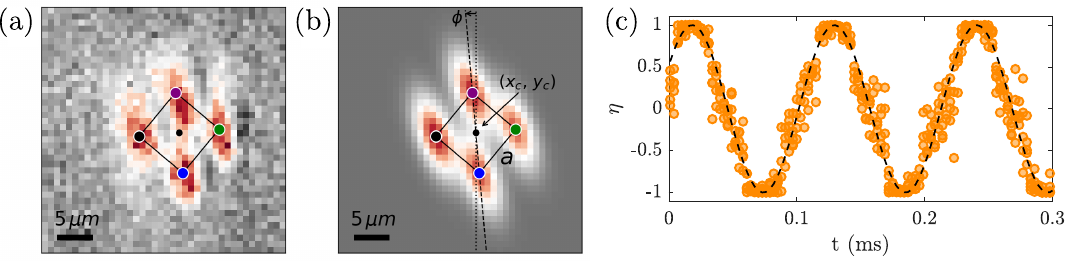}
		\caption{Exemplar fit procedure. a) Experimental image, b) fit result. The colored corners of the square, with edge length $a$, indicate the fitted positions of the four droplets. The dotted black line is the $y$-axis of the rotating reference frame, with the center in the guiding center origin ($x_c$, $y_c$) indicated by the solid black dot. The tilted dashed line measures the angular position $\phi$ of the purple droplet with respect to the $y$-axis, obtained from the fit. For this image, the extracted rotation is $\phi\,=\,-5.15^\circ$. (c) Cyclotron coordinate $\eta$ extracted from the fit of the experimental images with  (orange circles) and corresponding sinusoidal fit (black dashed line).}
		\label{fig:suppmat_analysis}
	\end{figure}

	To measure the angular position of the droplets from experimental images, we use a fit function. We first apply a Gaussian filter of size $\sigma\,=1$ px (${\simeq}\, 0.5 \,\mu\si{m}$) for noise reduction, before normalizing each image to the peak density. Each image is then fitted with a function defined as follows. Four elliptical 2D Gaussian density functions with four variable peak densities $n_j$, $j=0,1,2,3$ are centered on the corners of a square, with variable length $a$, phase $\phi$ and 2D center of mass position $x_c$, $y_c$ (guiding center origin), see Extended data Fig.\,\ref{fig:suppmat_analysis}, defining a total of 8 free parameters. The widths of the individual Gaussian distributions defining each droplet are not free parameters, but pre-calibrated to minimize fit residuals and fixed for all images. The orientation of each Gaussian (its cyclotron rotation) is also fixed and locked to the polarization angle, a feature we have verified for this dataset by first letting it as a free fit parameter, see Extended Data Fig.\,\ref{fig:suppmat_analysis}(c). This result is in agreement with previous works \cite{Klaus2022oov,casotti2024oov}. The angular position of the droplets is then given by $\phi$, assuming a circularly symmetric supersolid, such that $\phi_{\text{rot},j} = \phi + j\pi/2$. We probe the robustness of our fit by taking two unique initial conditions for each image. First, we set $\phi=0$ and perform the fit, then this is repeated for $\phi=\pi/4$, i.e.~the maximally different angle. We have verified that, independent of the initial condition, our results remain unchanged.
	
	\section{Fourier transform analysis}\label{ft} 
	To study the droplets' angular position in the rotating frame $\phi_{\mathrm{rot}}$ as a function of time, we performed a 2D Fourier transform from $(\phi_{\mathrm{rot}},t)$ to $(k_\phi,f)$ space of the diagrams in Fig.\,\ref{fig:fig3}(a)-(b), both before and after synchronization. For the experimental data, we first bin the coordinates $(\phi_{\mathrm{rot}},t)$ in a 2D histogram using a $120 \times 250$ grid and then apply the 2D Fourier transform for two time intervals, one before synchronization ($[0,50]$\,ms) and one after ($[60,110]$\,ms). The position of the peaks in Fourier space reflects the periodicity in time and space. The frequency $f$ corresponds to the droplets' rotation frequency in the rotating frame, divided by the number of droplets (four, in our case), and $k_\phi$ represents the angular periodicity of the droplets. Since the supersolid state consists of four droplets, spaced by $\pi/2$, all Fourier peaks have $k_\phi = 2/\pi \sim 0.6\,$rad$^{-1}$.
	In presence of Fourier peaks with a finite value of $k_\phi$ and $f$, the system is not synchronized, whereas when the frequency is peaked at $f=0$, the system is fully synchronized. We applied the same analysis to the theoretical data and observed a similar structure, with the peaks at finite $f$ being closer to zero, reflecting slower dynamics. 
	
\begin{figure}[h!]
		\includegraphics[width=\textwidth]{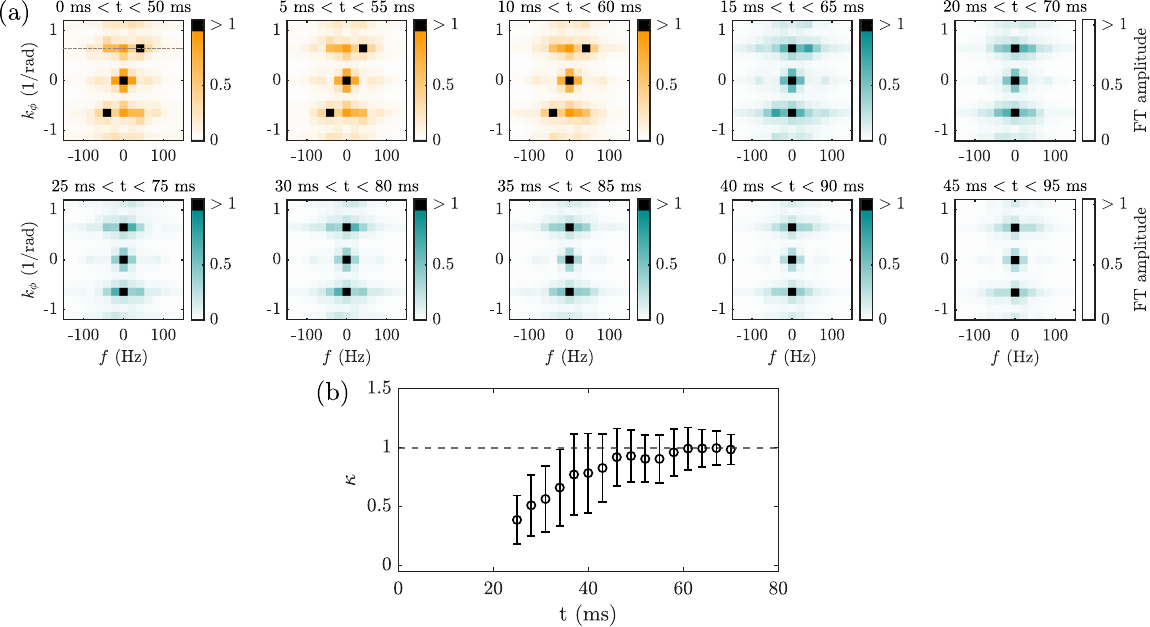}
		\caption{Extended analysis for the experimental data presented in Fig.\,\ref{fig:fig4} of the main text. (a) 2D Fourier transform of the experimental droplet angular position for varying time intervals. (b) Frequency alignment $\kappa$ as a function of time. The error bar is obtained from the FWHM of the Lorentzian function fit of the frequency components of the 2D Fourier transform along the dashed line shown in (a). This value is used as uncertainty for the guiding center frequency $\omega_g$, and propagated to determine the $\kappa$ uncertainty.}
		\label{fig:extended_fourier_analysis}
	\end{figure}

	In Extended Data Fig.\,\ref{fig:extended_fourier_analysis}(a) we show an extended version of the experimental analysis shown in the main text. The frequency position of the off-centered peaks divided by the number of droplets and converted in the laboratory frame is the experimental guiding center frequency $\omega_{g}$. We extract the frequency of the peak by fitting with a Lorentzian function the 2D Fourier transform along the dashed line, shown in the first subplot of (a). The error associated to this frequency is the FWHM of the Lorentzian function extracted from the fit. 
	Once obtained the $\omega_{g}$, we estimate the experimental frequency alignment $\kappa$, see Extended Data Fig.\,\ref{fig:extended_fourier_analysis}(b), where the experimental cyclotron frequency is fixed equal to the drive frequency, see Extended Data Fig.\,\ref{fig:suppmat_analysis}(c).
	
	\section{Numerical simulations of an $\Omega$ ramp}\label{ramp_omega}
\begin{figure}[h!]
		\includegraphics[width=0.7\columnwidth]{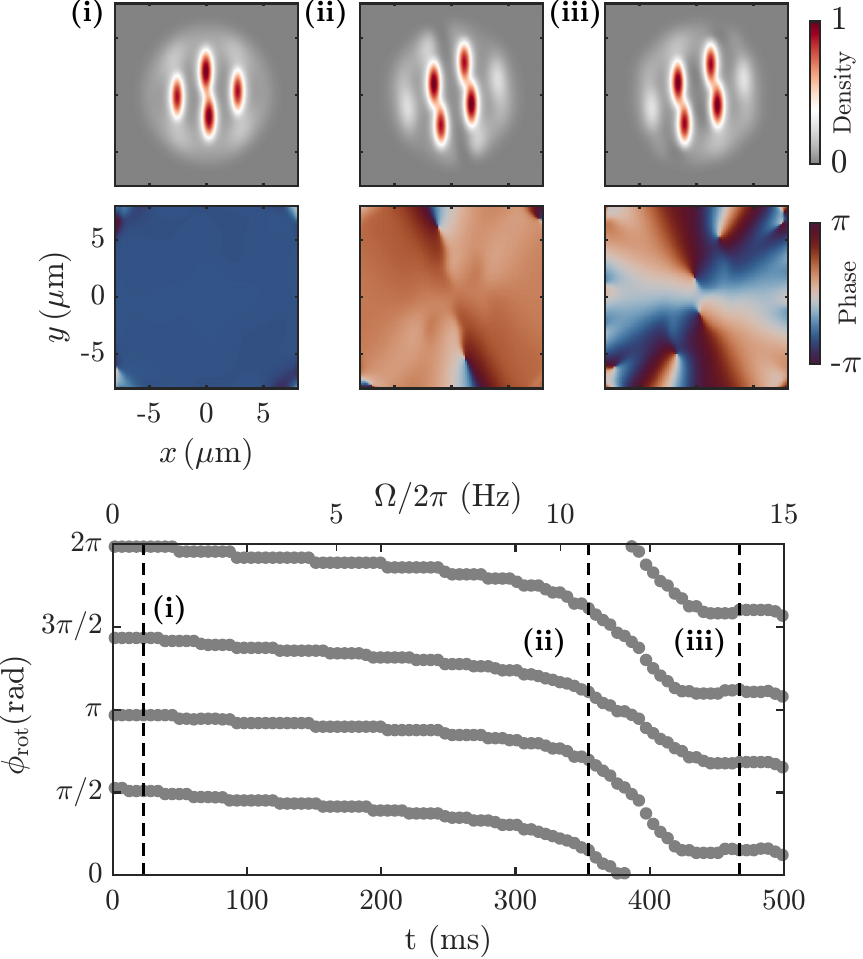}
		\caption{Numerical simulations for synchronization during rotation frequency ramp. The upper panels show the column density normalized to its maximum value and the central phase slice for some frames of the supersolid state during the simulation, selected at times (i)-(iii) indicated by the dashed vertical lines in the lower panel. The lower panel shows the  angular position of the droplets as a function of time, for a state obtained with parameters: $N=60000$, $a_s=90\,a_0$, $[\omega_x,\omega_y,\omega_z]=2\pi\times[51,50,149]$\,Hz, dissipation parameter $\gamma=0.05$.}
		\label{fig:ramp_theory}
	\end{figure}

	In the main text, we have shown that a slow increase of the rotation frequency leads to a desynchronization and subsequent resynchronization following a vortex nucleation. Here, we numerically simulate this situation with a protocol similar to what has been applied in the experiment, see Fig.\,\ref{fig:fig5}. In Extended data Fig.\,\ref{fig:ramp_theory} we show an example where the rotation frequency $\Omega/2\pi$ is linearly increased from 0 to 15 Hz over 500 ms. Throughout the full evolution, we track the center of mass of the droplets in the rotating frame, $\phi_{\mathrm{rot}}$. The theoretical simulation reveals the same behavior as the experimental data. Initially, the droplets remain mostly stationary in the rotating frame (i). As $\Omega$ exceeds the critical value for vortex nucleation, the droplets begin to lag behind the magnetic field (moving faster in the rotating frame), indicating a lack of synchronization. Around $t\sim350\,$ms the vortices approach the system (ii) and eventually enter, restoring the synchronization condition and reaching the stationary configuration in the rotating frame (iii). 
	
	\newpage
	\onecolumngrid
	
	\section{Extended data figures}

\end{document}